\documentclass[twocolumn,english,aps]{revtex4}
\usepackage[T1]{fontenc}
\usepackage[latin1]{inputenc}
\usepackage{amsmath}
\usepackage{graphicx}
\usepackage{setspace}
\usepackage{amssymb}

\makeatletter

\newcommand{\noun}[1]{\textsc{#1}}

\usepackage{babel}
\makeatother
\begin{document}

\preprint{This line only printed with preprint option}

\title{Optimal bounded-error strategies for projective measurements in non-orthogonal
state discrimination}

\author{M.A.P. Touzel}

\email{max.puelmatouzel@utoronto.ca}

\affiliation{Department of Physics}

\affiliation{Centre for Quantum Information and Quantum Control, University of
Toronto, 60 St. George St., Toronto, Canada M5S-1A7}

\author{R.B.A. Adamson}

\affiliation{Department of Physics}

\affiliation{Centre for Quantum Information and Quantum Control, University of
Toronto, 60 St. George St., Toronto, Canada M5S-1A7}

\author{A.M. Steinberg}

\affiliation{Department of Physics}

\affiliation{Centre for Quantum Information and Quantum Control, University of
Toronto, 60 St. George St., Toronto, Canada M5S-1A7}

\begin{abstract}
\begin{singlespace}
Research in non-orthogonal state discrimination has given rise to
two conventional optimal strategies: unambiguous discrimination (UD)
and minimum error (ME) discrimination. This paper explores the experimentally
relevant range of measurement strategies between the two, where the
rate of inconclusive results is minimized for a bounded-error rate.
We first provide some constraints on the problem that apply to generalized
measurements (POVMs). We then provide the theory for the optimal projective
measurement (PVM) in this range. Through analytical and numerical
results we investigate this family of projective, bounded-error strategies
and compare it to the POVM family as well as to experimental implementation
of UD using POVMs. We also discuss a possible application of these
bounded-error strategies to quantum key distribution.\end{singlespace}

\end{abstract}
\maketitle
PACS numbers: 03.65.Ta, 03.67.-a, 03.67.Dd, 03.67.Hk

\section{Introduction}

It is a well-known feature of quantum mechanics that it is impossible
to discriminate perfectly between non-orthogonal states. For example,
if a party is repeatedly sent one of two known, non-orthogonal states
and is asked each time, `Which of the two was sent?', the set of responses
based on measurements of the sent states must include either incorrect
or inconclusive responses or both. Incorrect responses occur when
the responding party misidentifies the state, while inconclusive ones
occur when the responding party replies that he doesn't know what
state was sent. The responding party knows only what the possible
states are and with what probability each is sent. This problem, called
quantum state discrimination, has played an important role in quantum
information science \cite{key-113}. There are two kinds of strategies
that are usually considered: the minimum error (ME) strategy and the
unambiguous discrimination (UD) strategy.

A strategy that minimizes the incorrect responses with no inconclusive
responses is known as the minimum error strategy. For two states,
it is obtained through a standard projection-valued measurement (PVM).
For $|\psi_{1}\rangle$ and $|\psi_{2}\rangle$, with respective prior
probabilities $\eta_{1}$ and $\eta_{2}=1-\eta_{1}$, the minimum
error rate as a function of the overlap between two states has an
analytic form given by Helstrom \cite{1} as \begin{equation}
P_{{\rm ME}}=\frac{1}{2}(1-\sqrt{1-4\eta_{1}\eta_{2}|\langle\psi_{1}|\psi_{2}\rangle|^{2}})\ .\label{eq:1}\end{equation}
When $\eta_{1}=\eta_{2}=1/2$, the PVM that achieves the minimum error
is oriented symmetrically around the states. By the orientation of
the measurement, we mean simply the orientation of the set of eigenstates
of the measurement in a vector representation of the Hilbert space.
We will discuss measurement elements in this way, i.e. in terms of
their eigenstate vectors.

A strategy that minimizes the inconclusive rate, $P_{{\rm In}}$,
with no incorrect responses is known as the unambiguous discrimination
strategy. The absence of errors allows for conclusive, i.e. certain,
discrimination; $P_{{\rm Con}}=1-P_{{\rm In}}$ is called the conclusive
rate and the UD problem is often phrased as maximizing $P_{{\rm Con}}$.
For two states, after setting $\eta_{1}\geq\eta_{2}$ without loss
of generality, the maximum conclusive rate for projective measurements
is \begin{equation}
P_{{\rm Con}}^{{\rm PVM}}=\eta_{1}(1-|\langle\psi_{1}|\psi_{2}\rangle|^{2}).\label{eq:3}\end{equation}
However, a generalized or positive operator-valued measurement (POVM)
is in fact optimal under the condition that $(\eta_{2}/\eta_{1})^{1/2}|\langle\psi_{1}|\psi_{2}\rangle|\geq|\langle\psi_{1}|\psi_{2}\rangle|^{2}$
\cite{key-112}. Originally addressed by Ivanovic, Dieks, and Peres
(IDP) \cite{key-99}, the measurement gives the optimal conclusive
rate \cite{key-112}\begin{equation}
P_{{\rm Con}}^{{\rm POVM}}=1-2(\eta_{1}\eta_{2})^{1/2}|\langle\psi_{1}|\psi_{2}\rangle|.\label{eq:2}\end{equation}
Unambiguous discrimination strategies are central to quantum key distribution
(QKD) in quantum cryptographic protocols \cite{key-102,key-95} and,
thus, the success rate of the protocol is dependent on what type of
measurement, i.e. a PVM or POVM, one chooses to implement. In the
case of equal priors, $\eta_{1}=\eta_{2}=1/2$, a POVM is the optimal
measurement for any overlap. For example, when the overlap is $1/\sqrt{2}$
, the optimal POVM performing UD gives a maximum conclusive rate of
29.3\%, whereas the optimal PVM gives a maximum conclusive rate of
25\%. The difference between the results for PVMs and POVMs is more
pronounced in a particular three-state example of UD from \cite{key-101},
which we discuss later on, where the conclusive rate of the optimal
POVM is more than twice that given by the corresponding optimal PVM. 

The advantage that POVMs provide over PVMs in UD is related to the
fact that POVM elements are \emph{not} restricted to being orthogonal.
Thus, their number can exceed the dimension of the system's Hilbert
space (which we will assume throughout equals the number of input
states). A positive operator is associated with each state and one
extra operator is defined for inconclusive results. In contrast, PVM
elements \emph{are} mutually orthogonal. Since unambiguously discriminating
a particular input state, $|\psi_{i}\rangle$, involves knowing for
certain that the sent state was not any of the other input states,
the respective PVM element is oriented orthogonally to all other input
states so that if the outcome corresponding to that element is obtained,
one knows for certain that the sent state was $|\psi_{i}\rangle$.
Only one such element exists in general for PVMs and thus only one
input state may be unambiguously discriminated. In this case, the
projector for the orthogonal subspace corresponds to inconclusive
results. Ultimately, however, if the input states do not form a linearly
independent set, for example when the number of input states is larger
than the dimension of the Hilbert space, UD is not possible in general,
regardless of the type of measurement \cite{key-110}. 

In realistic quantum information processing, noisy channels are inevitable,
even in UD, so a more general class of strategies where the responding
party gives both inconclusive \emph{and} incorrect responses becomes
useful. Since all outcomes are now error-prone, conclusive results
no longer exist. For this case, an approach that maximizes the correct
rate of individual outcomes has been given in \cite{key-115}. This
so-called `maximum confidence' measurement gives the highest probability
that the given interpretation of a result was correct. In this paper,
we instead adopt the equivalent approach of minimizing the inconclusive
rate, given some bounded-error rate, since we wish to consider error
as a fixed parameter of the problem. 

In either approach, UD and ME schemes exist as limiting cases: the
latter when the inconclusive rate is 0 and the former when the error
rate is 0. In between, trade-offs exist between the two rates. In
the two-state case, for example, one may achieve strategies that at
once give error rates less than the Helstrom bound and inconclusive
rates less than the IDP bound. Zhang, Li, and Guo derive in \cite{key-103}
a general inequality for this intermediate range for the two input
states, $|\psi_{1}\rangle$ and $|\psi_{2}\rangle$, given as $P_{{\rm In1}}P_{{\rm In2}}\geq|P_{{\rm IP}}-\sqrt{P_{{\rm C1}}P_{{\rm E2}}}-\sqrt{P_{{\rm C2}}P_{{\rm E1}}}|^{2}$,
where $P_{{\rm In1}}$ and $P_{{\rm In2}}$ are the inconclusive rates,
$P_{{\rm C1}}$ and $P_{{\rm C2}}$ are the correct rates, and $P_{{\rm E1}}$
and $P_{{\rm E2}}$ are the error rates, respectively, and $P_{{\rm IP}}=\langle\psi_{1}|\psi_{2}\rangle^{M}$
for the slightly more general problem of sending $M$ copies of either
$|\psi_{1}\rangle$ or $|\psi_{2}\rangle$ each time. In the case
that each state is equally likely ($\eta_{1}=\eta_{2}$) and $M=1$,
the inequality reduces to one given in an early work on this intermediate
range published by Barnett and Chefles \cite{key-89}. A few works
\cite{key-93,key-97} have extended the problem to mixed states where
the possibility of UD is conditional \cite{key-114} because of a
possible linear dependency between the states. The minimum error rate
for the problem is thus not necessarily zero and the intermediate
range of bounded-error strategies becomes necessary. In \cite{key-97},
Eldar formulates the problem in terms of semi-definite programming,
a branch of convex optimization. Once formulated in this way, powerful
numerical techniques can be applied that readily give the optimal
inconclusive rate. Applying these techniques, Eldar shows that the
conditions for the optimal mixed state solution provided in \cite{key-93}
are necessary and sufficient; i.e. they guarantee a global optimum.
These techniques scale efficiently with dimension and so are used
in this work to access the more analytically difficult, higher dimensional
problems in quantum state discrimination using POVMs.

Beyond their theoretical interest, the intermediate range of strategies
in quantum discrimination is not only important in accounting for
noise, but because the introduction of error can actually be beneficial.
Specifically in ideal schemes, introducing error to an UD strategy
allows for more correct discriminations (though, of course, by sacrificing
any truly unambiguous response). Both the receiver and an eavesdropper
can most likely make use of this fact in QKD \cite{key-93}. For example,
a recent counter-intuitive result shows that the information attainable
by an eavesdropper in QKD can decrease with the introduction of error
\cite{key-104}. 

Even though POVMs obtain the optimal solution in general, the projective
versions of these bounded-error strategies deserve study since PVMs
are still widely used in practice and the differences between the
two types of measurement in this context have yet to be studied.

In the following section, we establish some constraints on the optimal
solution to the problem of allowing for both inconclusive and incorrect
responses. These constraints are then used in phrasing the problem
of minimizing the inconclusive rate, for projective strategies, given
a bounded-error rate. An instructive example of a two-state PVM solution
is then presented in Section 3, after which higher dimensional problems
are discussed in Section 4 with an example of a three-state case based
on a previous experiment \cite{mycitation}. Consequently, a claim
made in \cite{mycitation} regarding the superiority of POVMS over
PVMs in UD is amended. Lastly, by applying the bounded-error strategy
to the B92 protocol, we discuss increasing the key generation rate
and summarize our results in Section 5.

\section{Problem Formulation}

For the $n$-state problem, given any set of $n$ input states, $|\psi_{1}\rangle$,$|\psi_{2}\rangle$,
..., $|\psi_{n}\rangle$, with prior probabilities $\eta_{1}$, $\eta_{2}$,
..., $\eta_{n}$, respectively, we must find the function, $\tilde{P}_{\textrm{In}}(\epsilon)$,
representing the minimum inconclusive response rate as a function
of a bounded-error rate, $\epsilon$, defined as the largest tolerable
fraction of incorrect responses out of the total number of sent states.
In other words, the responding party is allowed a maximum average
number of incorrect responses and, under this restriction, tries to
minimize the average number of inconclusive responses. 

Given an existing, but not necessarily optimal strategy with some
particular inconclusive rate, $P_{{\rm In_{0}}}$, and some particular
error rate, $P_{{\rm E_{0}}}$, consider the following two manipulations
of that strategy that decrease the error rate and inconclusive rate,
respectively, and provide some constraints on the problem. (Refer
to Fig.1.)

First, the error rate, $P_{{\rm E}}$, may be decreased from $P_{{\rm E_{0}}}$
by randomly calling inconclusive some fraction of the results obtained
from error-prone outcomes (those that we interpret as a particular
sent state but that are sometimes incorrect). Thus, in Fig.1 there
is a family of strategies represented by the line segment on the plot
connecting the point $(P_{{\rm E_{0}}},P_{{\rm In_{0}}})$ and the
point $(0,\ 1)$ at the upper left-hand corner. Moving from right
to left along this line an increasing fraction of responses are inconclusive.
When all of the responses are inconclusive the error rate is naturally
zero. This property places a constraint on the values of $\tilde{P}_{\textrm{In}}(\epsilon)$
at neighboring points. %
\begin{figure}
\includegraphics{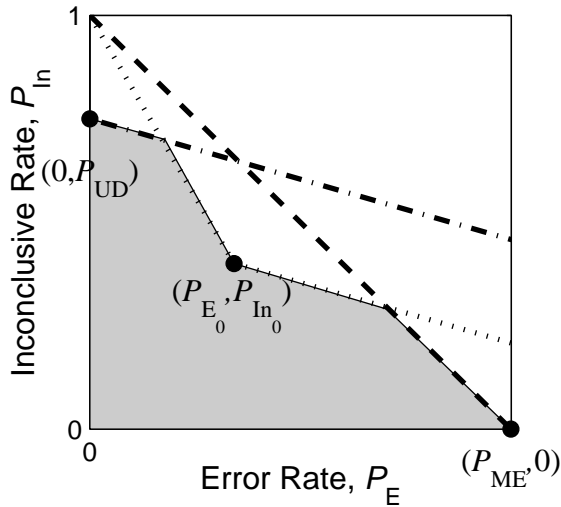}

\caption{Bounds on the minimum inconclusive rate function, $\tilde{P}_{\textrm{In}}(\epsilon)$,
for the $n$-state problem with equal prior probabilities. The dash-dotted
line with a slope of $-\frac{n}{n-1}$ starting at $(0,P_{{\rm UD}})$
represents strategies where random guesses are made when inconclusive
outcomes are obtained from the UD strategy. The dashed line that connects
the point $(0,1)$ to $(P_{{\rm ME}},0)$ represents strategies where
results for error-prone outcomes from the ME strategy are interpreted
as inconclusive. Similar strategy manipulations (shown as dotted lines)
apply to any existing strategy, represented by $(P_{{\rm E_{0}}},P_{{\rm In_{0}}})$,
and so provide two constraints on the minimizing function, $\tilde{P}_{\textrm{In}}(\epsilon)$:
a bound on its slope (it must be less than$-\frac{n}{n-1}$) and a
bound on the neighboring points of the function. $\tilde{P}_{\textrm{In}}(\epsilon)$
then lies somewhere in the shaded region.}
\end{figure}
Second, the inconclusive rate, $P_{\textrm{In}}$ may be decreased
from $P_{{\rm In_{0}}}$ by randomly guessing the sent state any fraction
of the time an inconclusive outcome is obtained. For example, in the
case of equal prior probabilities, a random guess will be correct
with probability $1/n$ and incorrect with a probability of $1-1/n$
. The error rate is therefore increased above $P_{{\rm E_{0}}}$ by
$1-1/n$ of the rate of guessed outcomes. Thus, the line with slope
$-\frac{n}{n-1}$ beginning at and to the right of $(P_{{\rm E_{0}}},\  P_{{\rm In_{0}}})$
contains strategies that differ only in the fraction of the time an
inconclusive outcome is replaced by a randomly guessed outcome (See
Fig.1). For a given error rate, the optimal strategy must have an
inconclusive rate less than or equal to the value along this line.

Applied to the ME and UD strategies, the above two strategy manipulations
restrict the optimal solution so that it must lie at or below the
line from $(0,\ 1)$ to the ME point, $(P_{{\rm ME}},0)$, and at
or below the line from the UD point, $(0,P_{{\rm UD}})$, to $(\frac{n}{n-1}(1-P_{{\rm UD}}),0)$,
which is the right-endpoint of a line with slope $-\frac{n}{n-1}$
beginning at $(0,P_{{\rm UD}})$ (See Fig.1). The constraints displayed
in Fig.1 are satisfied by analytical results given in \cite{key-89}
for two states using POVMs and presumably for all POVM solutions.
We now restrict ourselves to projective strategies. 

An $n$-element PVM is a set of orthogonal projectors, $\{{\rm P_{1}},{\rm P}_{2},\ ...\ ,\ {\rm P}_{n}\}$,
where ${\rm P}_{i}=|p_{i}\rangle\langle p_{i}|$ for some vector $|p_{i}\rangle\in\mathbb{C}^{n}$
and $\langle p_{i}|p_{j}\rangle=\delta_{ij}$, for $i,\  j=1,\ ...,\  n$.
We will continue to discuss the PVM elements in terms of these vectors
onto which the elements project. 

Recall that in the ME strategy, each ${\rm P}_{i}$ corresponds to
a possible measurement outcome, which is interpreted (occasionally
incorrectly) as the particular input state, $|\psi_{i}\rangle$, having
been sent. We now formalize the idea of only making that interpretation
a fraction of the time that we obtain that particular outcome; the
rest of the time we call the result inconclusive. 

Consider the following strategy. To each ${\rm P}_{i}$ associate
a fraction, $w_{i}$, called the discrimination weight of ${\rm P}_{i}$.
When the projective measurement is performed many times, the outcome
corresponding to ${\rm P}_{i}$ will be obtained many times and $w_{i}$
is the fraction of those outcomes that should be interpreted as $|\psi_{i}\rangle$.
The rest should be interpreted as inconclusive. A `discriminated'
input state or measurement element is defined as one with $w_{i}=1$.
In the ME strategy, all states are discriminated input states whereas
in the UD strategy only one input state is discriminated.

Taking these discrimination weights into account, the expression for
the correct rate, $P_{{\rm C}}$, is

\begin{equation}
P_{{\rm C}}=\sum_{i=1}^{n}w_{i}\eta_{i}|\langle p_{i}|\psi_{i}\rangle|^{2}\label{eq:4}\end{equation}
 and the expression for the error rate, $P_{E}$, is \begin{equation}
\  P_{{\rm E}}=\sum_{i=1,\  j=1,\  i\neq j}^{n}w_{i}\eta_{j}|\langle p_{i}|\psi_{j}\rangle|^{2}.\label{eq:5}\end{equation}
The inconclusive rate is then defined as $P_{{\rm In}}=1-(P_{{\rm C}}+P_{{\rm E}})$
and the optimization problem for some bounded-error rate, $\epsilon$,
is:\begin{equation}
\begin{array}{cc}
\textrm{minimize} & P_{{\rm In}}=1-(P_{{\rm C}}+P_{{\rm E}})\\
\textrm{\  subject\  to} & P_{{\rm E}}\leq\epsilon\ ,\ \end{array}\label{eq:6}\end{equation}
 over the discrimination weights $w_{1}$, ..., $w_{n}$, and the
orientation of the PVM through $p_{1}$, ..., $p_{n}$. The minimal
function, $\tilde{P}_{\textrm{In}}(\epsilon)$, is obtained by varying
$\epsilon$ from $0$ to $P_{{\rm ME}}$. On a plot of $P_{{\rm In}}$
versus $P_{{\rm E}}$, $\tilde{P}_{\textrm{In}}(\epsilon)$ is a curve.
The region above and to the right of that curve contains points representing
inconclusive rate and error rate pairs that can be obtained with a
PVM strategy. At one endpoint of the curve is the UD strategy at $\epsilon=0$,
where all but one of the discrimination weights are constrained to
be zero since at most one input state may be unambiguously discriminated.
At the other end is the ME strategy at $P_{{\rm In}}=0$ where a discrimination
is attempted for all states and so all discrimination weights are
equal to 1. With increasing error from 0, $\tilde{P}_{\textrm{In}}(\epsilon)$
may be obtained by optimally increasing the values of the $n-1$ weights
from 0 to 1 while adjusting the orientation of the PVM. Now that the
problem is formulated, we proceed with some examples.

\section{A Two-State Example}

Consider the restricted problem where there are two states to be discriminated
and all their coefficients are real. This case has a simple and instructive
geometrical interpretation where the PVM is represented by a set of
two orthogonal vectors $p_{1}$ and $p_{2}$ on the unit circle in
$\mathbb{R}^{2}$ whose corresponding outcomes are interpreted as
the states $|\psi_{1}\rangle$ and $|\psi_{2}\rangle$, respectively
(See Fig. 2). We now focus on extending the two known strategies (ME
and UD) into the intermediate range.

In the UD strategy where $P_{{\rm E}}=0$ and assuming again without
loss of generality that $\eta_{1}\ge\eta_{2}$, $\psi_{1}$ is discriminated
using the outcome corresponding to $p_{1}$ by setting $w_{1}=1$,
$w_{2}=0$ and $|\langle p_{1}|\psi_{2}\rangle|=0$. The correct rate
is then \begin{equation}
P_{{\rm {\rm C}}}=\eta_{1}|\langle p_{1}|\psi_{1}\rangle|^{2}.\label{eq:7}\end{equation}
 A discrimination is not attempted for $|\psi_{2}\rangle$, leaving
the outcomes for $p_{2}$ as inconclusive. Without changing the discrimination
weights, this inconclusive rate may be decreased via a rotation of
the PVM from the UD orientation so that the overlap between $p_{2}$
and both $\psi_{1}$ and $\psi_{2}$ decreases. In the process, the
overlap between the $p_{1}$ and $\psi_{1}$ increases thereby increasing
the correct rate. Error is introduced in the process since there is
now a non-zero overlap between the discriminated PVM element, $p_{1}$,
and the input state to which it is not associated, $\psi_{2}$. This
error rate is\begin{equation}
P_{{\rm E}}=\eta_{2}|\langle p_{1}|\psi_{2}\rangle|^{2}.\label{eq:8}\end{equation}
The increase in the two correct and error rates can be found using
the setup in Fig.2. Equations (7) and (8) become\begin{equation}
P_{{\rm C}}=\eta_{1}\sin^{2}(\phi+\theta)\ ,\ \ \ \  P_{{\rm E}}=\eta_{2}\sin^{2}(\phi),\label{eq:9}\end{equation}
respectively, where $\phi$ is the angle of rotation that is set by
the parameter $P_{{\rm E}}$. For small $\phi$, and $\theta$ near
$\pi/4$, the correct rate in (9) increases linearly and the error
rate increases quadratically. Thus, near $\phi=0$ we can obtain a
significant increase in the correct response rate without a correspondingly
large increase in the error rate (providing a maximum benefit when
$\theta=\pi/4$).

As $\phi$ increases with increasing $P_{{\rm E}}$, a continuous
set of pairs of inconclusive and error rate values are achieved through
this freedom in the orientation of the measurement. At the curve's
left-endpoint, $P_{{\rm E}}=0$ so that $\phi=0$ by Eq. (9) and we
regain Eq. (3) as $P_{{\rm UD}}^{{\rm PVM}}=\eta_{1}(1-{\rm cos}^{2}(\theta)).$
\begin{figure}
\includegraphics{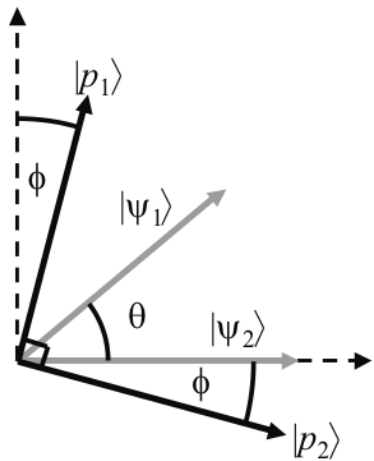}

\caption{The geometry of the real, two-state problem. The input states are
represented by $|\psi_{1}\rangle$ and $|\psi_{2}\rangle$, and the
PVM elements project onto $p_{1}$ and $p_{2}$. The angle $\phi$
offsets the orientation of the PVM from the zero-error UD case where
$p_{1}$ is interpreted as $|\psi_{1}\rangle$ and $p_{2}$ is interpreted
as inconclusive. $\theta$ is the separation angle between the states.}
\end{figure}
 The expression for this curve corresponding to the correct rate in
(9) as a function of the increasing error rate, $P_{E}$, is

\begin{equation}
P_{{\rm C}}=P_{{\rm UD}}+\frac{\eta_{1}}{\eta_{2}}P_{{\rm E}}\left(\textrm{cos}(2\theta)+\sqrt{{\frac{\eta_{2}}{P_{{\rm E}}}-1}}\ \ \textrm{sin}(2\theta)\right).\label{eq:8}\end{equation}
When taking the limit $P_{{\rm E}}\to0$, $P_{{\rm C}}\propto\sqrt{P_{{\rm E}}}$
and so the slope of the curve given by (10) becomes infinite. Again,
there is a substantial increase (decrease) in the correct (inconclusive)
rate with a small increase in error. An example of this `$w_{2}=0$'
curve is shown in Fig.3 along with other results discussed below for
the case of $\eta_{1}=\eta_{2}$ and $\theta=\frac{\pi}{4}$ .

Focusing now on the ME strategy, error may be decreased from $P_{{\rm ME}}$
by making use of the freedom of the discrimination weights, i.e by
calling error-prone outcomes inconclusive. This is accomplished by
decreasing either of the discrimination weights, $w_{1}$ or $w_{2}$,
from 1. Taking one of them to 0 produces a linear curve on the plot
in Fig. 3, shown for the case of $\eta_{1}=\eta_{2}$ and $\theta=\frac{\pi}{4}$.

The inconclusive rate function for the optimal strategy, $\tilde{P}_{{\rm In}}(\epsilon)$,
is obtained by rotating the PVM while \emph{simultaneously} changing
the discrimination weights. The optimal strategy smoothly changes
from the orientation-dependent strategy at errors near 0 to the weight-dependent
strategy at errors near $P_{ME}$. For intermediate errors, both strategies
are significant and are jointly used to achieve the optimum. The numerical
solution for $\theta=\frac{\pi}{4}$ and $\eta_{1}=\eta_{2}$ is shown
in Fig.3 where a maximum 10\% increase in the correct discrimination
rate is achieved by both rotating the PVM \emph{and} changing the
discrimination weights as compared to doing either one by itself.

Also shown in Fig.3 is the corresponding optimal POVM curve as given
in \cite{key-89}. It always performs better than the PVM solution,
as it must. However, even though the curves representing the two strategies
are diverging as $P_{E}$ increases from 0, this trend is short-lived.
With increasing error, the gap between their inconclusive rates is
made smaller by the use of the discrimination weight, $w_{2}$, in
the optimal PVM strategy until the two solutions finally converge
on each other at the ME values. %
\begin{figure}
\includegraphics{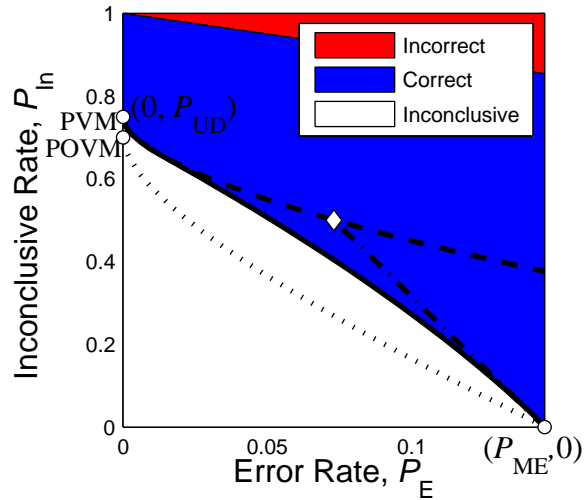}

\caption{(color on-line). The two-state discrimination problem for $\theta=\pi/4$
and $\eta_{1}=\eta_{2}=1/2$. As $P_{{\rm E}}\to0$, the optimal PVM
curve (solid line) approaches the dashed curve representing the less
general optimization over the orientation of the PVM with $w_{2}=0$.
For error approaching the ME value, the optimal PVM curve approaches
the dash-dotted line representing strategies based on the ME strategy
but with $0<w_{2}<1$. For any value of $\theta$, the dashed and
dash-dotted curves intersect at $(\frac{P_{{\rm ME}}}{2},\frac{1}{2})$
where $w_{2}=0$, shown here as a diamond marker. The optimal POVM
curve for the same parameters is shown as the dotted line.}
\end{figure}

\section{Higher Dimensional Problems and an Example}

The underlying structure of the optimal strategy given above is not
restricted to two states, but rather generalizes to problems with
a larger number of input states. In general for $n$ input states
there will be $n$ curves analogous to that given by (10), where each
represents a family of strategies, denoted ${\rm C}_{m}$, for $1\leq m\leq n$,
with a fixed, discrete number of discriminated ($w_{i}=1$) input
states. Each ${\rm C}_{m}$ is defined by the discrimination weights:
$w_{i}=1,\ \textrm{for}\  i=1,\ ...,\  m$, and $w_{i}=0$ otherwise.
${\rm C}{}_{m}$ represents the set of strategies where one always
tries to discriminate the optimal subset of $m$ input states with
$m$ projectors and interprets the remaining $n-m$ outcomes as inconclusive.
For a given set of discriminated states, i.e. for a given ${\rm C}_{m}$,
the error rate can be changed by reorienting the PVM, altering the
overlap between the input states and PVM elements. An optimal reorientation
generates, for each ${\rm C}_{m}$, a continuous curve in the plane
of $P_{{\rm In}}$ versus $P_{{\rm E}}$. Each ${\rm C}_{m}$ has
a minimum error rate, $P_{{\rm E},m}^{{\rm min}}$, that is non-zero
for $m>1$. All strategies that discriminate $m$ states must have
an error rate of at least $P_{{\rm E},m}^{{\rm min}}$. To obtain
a lower error rate $m$ must be reduced. For ${\rm C}_{1}$, the minimum
error is 0, i.e. in the UD case. For ${\rm C}_{n}$ the minimum error
is that of the ME case, $P_{{\rm ME}}$, where all weights equal 1
and ${\rm C}_{n}$ is in fact just the point $(P_{{\rm ME}},0)$.
The intermediate set of strategies for $1<m<n$ have a minimum error
lying between 0 and $P_{{\rm ME}}$, with $P_{{\rm E},m}^{{\rm min}}<P_{{\rm E},m+1}^{{\rm min}}$.

Naively, one could obtain a suboptimal set of discrimination strategies
as a function of allowed error by adopting the appropriate strategies
in ${\rm C}_{m}$ (by optimally reorienting the PVM) once $P_{{\rm E}}\geq P_{{\rm E},m}^{{\rm min}}$
and until $P_{{\rm E}}\geq P_{{\rm E},m+1}^{{\rm min}}$. However,
the \emph{optimal} strategy also involves the continuous transformation
of the discrimination weights along with the PVM reorientation and
generates a continuous, smooth minimum inconclusive rate curve between
0 and $P_{{\rm ME}}$ error. 

For any ${\rm C}_{m}$, there will be one of the $m$ discriminated
states (up to symmetries in that optimal subset) that will give the
worst contribution to the error. Reducing the weight for this state
gives the minimal increase in the inconclusive rate for errors below
$P_{{\rm E},m}^{{\rm min}}$ down to $P_{{\rm E},m-1}^{{\rm min}}$
at which point one is forced to reduce a weight of the remaining $m-1$
states to get the error any lower. Just as in the two-state case,
the optimal strategy is a smooth transition from the orientation-focused
strategies at errors just above $P_{{\rm E},m-1}^{{\rm min}}$ to
the weight-focused strategies at errors just below $P_{{\rm E},m}^{{\rm min}}$.
The latter can be seen in the dashed-dotted lines in figs. 3 and 4
where the weight of the respective worst state is reduced without
changing the orientation of the optimal PVM at the respective $P_{{\rm E},m}^{{\rm min}}$
value. Intuitively, the `worst' state is hardest to discriminate because
it is has the largest overlap with the rest of the states. More technically,
it has the largest ratio of error rate to correct rate contribution,
so reducing its discrimination weight gives the minimal increase in
the inconclusive rate for a given reduction in error rate.

The optimal curve can be found numerically by solving (6) over the
discrimination weights and the orientation of the PVM. The optimal
projective strategies arising from the above construction may be used
in a comparison with optimal POVM strategies for higher dimensional
problems, an example of which follows in the next paragraph.

What spurred interest into bounded-error projective strategies was
an earlier paper from our group \cite{mycitation} in which an optical
realization of UD was performed on a particular triplet of states
using the optimal POVM as suggested in \cite{key-101}. After a comparison
with the theoretical result of the corresponding optimal UD PVM, the
claim was made that this POVM had demonstrated {}``an improvement
of more than a factor of 2 over \emph{any} possible projective measurement''
(our emphasis). The measurement was of course accompanied by some
experimental error (3\% in this case, with a 2\% decrease in the inconclusive
rate as a result). Therefore, the legitimate comparison is between
the implemented POVM and the optimal PVM, implemented or theoretical,
that also gives that error rate. The family of projective strategies
for a bounded error given in this paper contain such an optimal projective
measurement. It is true that in any implementation these optimal projective
strategies would themselves acquire experimental errors that would
most likely make them less effective than the corresponding implemented
UD POVM. However, it is the general claim regarding realistic POVMs
and any PVM that we wish to address. Therefore, should any of these
theoretically optimal PVM measurements perform better than the implemented
POVM, the claim made in \cite{mycitation} would be invalidated. The
proper comparison is shown in Fig. 4 and it is clear that the optimal
PVM strategy that is wrong 3\% of the time answers correctly more
often than the implemented POVM in \cite{mycitation}, represented
at the 3\% error it achieved in the experiment. The optimal PVM elements
${\rm P_{1}}$, ${\rm P}_{2}$, and ${\rm P}_{3}$, given by the vectors
$p_{1}=(-0.63,0.63,0.45)$, $p_{2}=(0.71,0.71,0)$, and $p_{3}=(0.31,-0.31,0.90)$,
achieve a 62.3\% correct response rate as compared to the 54.5\% given
by the implemented POVM. 

Also shown in Fig.4 is the optimal curve for \emph{any} POVM, i.e.
the one with no experimental error, found with the duality techniques
described in \cite{key-97} using the program YALMIP. The large advantage
that the POVM solution has over the PVM solution in UD diminishes
quickly with increasing error from 0 because of the ability of PVMs
to make an increasingly large number of correct discriminations of
a second state while introducing very little error. A logarithmic
scale was used in Fig.4 to display this fact more clearly. For example
in Fig.4, the 25.4\% correct discriminations for PVMs at 0 error jumps
to over 50\% at 0.0025\% error. By contrast, for the POVM, the 54.5\%
correct discrimination rate at 0 error only goes up to 61.5\% by 0.0025\%
error. Experimentally accessing these regions of near-zero error where
POVMs give a significant advantage may prove difficult. Also, we note
that the effect is even stronger in Fig.4 than in the two-state case
considered above in Fig. 3. It therefore may be true that in experiments
using a large number of input states and those for which the experimental
error can not be made small, POVMs cease to give a sufficient advantage
over PVMs to warrant the increased practical difficulties in their
implementation.

\begin{figure}
\includegraphics{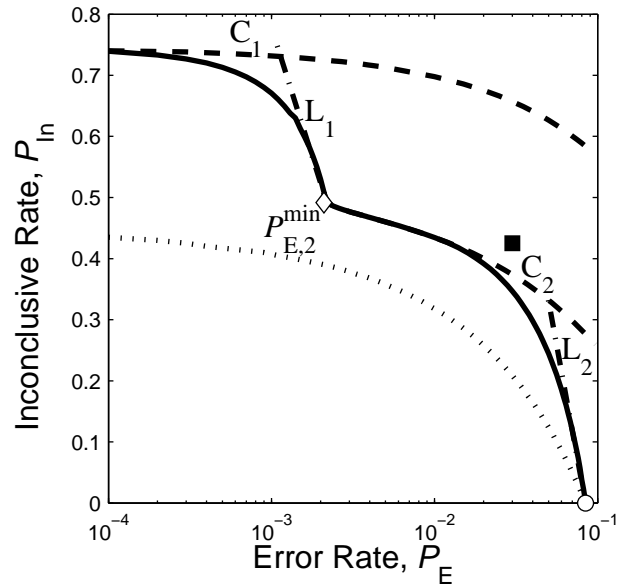}

\caption{State discrimination for the three states used in \cite{mycitation},
$|\psi_{1}\rangle=(\sqrt{2/3},\,0,\,1/\sqrt{3})$, $|\psi_{2}\rangle=(0,\,1/\sqrt{3},\,\sqrt{2/3})$
and $|\psi_{3}\rangle=(0,-1/\sqrt{3},\,0,\,\sqrt{2/3})$. The optimal
projective strategies are represented by the solid line. The two dashed
lines, ${\rm C_{1}}$ and ${\rm C}_{2}$, represent the strategies
that discriminate one and two input states, respectively, using only
the orientation of the PVM. The diamond and circular marker represent
the minimum error for ${\rm C}_{2}$, $P_{{\rm E,2}}^{{\rm min}}$,
and the minimum error, $(0,P_{{\rm ME}})$, respectively. Using the
PVM orientation defined at those points, ${\rm L}_{1}$and ${\rm L_{2}}$
are generated by reducing the weight corresponding to the discriminated
state ($w{}_{i}=1$) that is hardest to discriminate. The optimal
PVM at 3\% error gives a correct rate of 62.3\% compared to the 54.5\%
attained by the experimental POVM in \cite{mycitation} shown here
as a square. The optimal POVM is shown as the dotted line.}
\end{figure}

\section{Discussion and Summary}

One application for optimal bounded-error strategies may lie in increasing
quantum key generation rates. For example in the B92 protocol \cite{key-95},
Alice and Bob use the transmission and a UD measurement of two non-orthogonal
states in different bases to build a cryptographic key. The key generation
rate is $R=N(1-f_{{\rm In}})(1-H(e_{{\rm b}})-H(e_{{\rm p}}))$, where
$N$ stands for the number of sent states measured in the same basis
and $f_{{\rm In}}$ refers to the minimum inconclusive rate for the
measurement, found through, for example, the convex methods mentioned
in the introduction. $f_{{\rm In}}$ is dependent only on the fixed
separation between the two states and their prior \emph{}probabilities.
Out of all the counts that pass the procedure, $H(e_{{\rm b}})$ is
the fraction that are sacrificed to find the quantum bit-error rate
and $H(e_{{\rm p}})$ is the fraction lost in the privacy amplification
process. As long as $H(e_{{\rm b}})$ and $H(e_{{\rm p}})$ remain
within the bounds required for security, Bob is free to select a measurement
that offers him the highest key generation rate through its effect
on $f_{{\rm In}}$. A bounded-error strategy will perform better in
this regard than the UD measurement that is normally used. An unconditional
security proof for B92 is given in \cite{key-106} and provides estimates
for the bounds of $H(e_{{\rm b}})$ and $H(e_{{\rm p}})$. One would
need only to adjust the proof by considering a bounded-error strategy
instead.

In this work we have developed the general problem of the optimal
PVM that interpolates between the UD and ME strategies by minimizing
the inconclusive rate for some bounded-error rate. This range of strategies
is more experimentally relevant since errors are inevitable and our
choice of measurement is relevant since PVMs are more widely used
than POVMs. We have found, in both two and three-state examples, that
a small introduction of error leads to a large decrease in the inconclusive
rate for PVMs, which suggests that the substantial difference in UD
results for PVMs and POVMs may not exist once realistic errors are
considered.

\begin{acknowledgments}
This research was supported by NSERC and CIAR. The authors would like
to thank Janos Bergou and Masoud Mohseni for useful discussions. M.P.
Touzel also wishes to thank Peter Turner for some initial POVM Matlab
code and Fred Fung for many digressions relating to quantum information. 
\end{acknowledgments}

\end{document}